# Dynamic causal modelling of immune heterogeneity

Thomas Parr[1*], Anjali Bhat[1], Peter Zeidman[1], Aimee Goel[2], Alexander J. Billig[3], Rosalyn Moran[4], Karl J. Friston[1]

[1] Wellcome Centre for Human Neuroimaging, Queen Square Institute of Neurology, UK
[2] Royal Stoke University Hospital, UK
[3] UCL Ear Institute, University College London, UK
[4] Centre for Neuroimaging Science, Department of Neuroimaging, IoPPN, King's College London, UK

[*] Correspondence: thomas.parr.12@ucl.ac.uk

## Abstract

An interesting inference drawn by some Covid-19 epidemiological models is that there exists a proportion of the population who are not susceptible to infection – even at the start of the current pandemic. This paper introduces a model of the immune response to a virus. This is based upon the same sort of mean-field dynamics as used in epidemiology. However, in place of the location, clinical status, and other attributes of people in an epidemiological model, we consider the state of a virus, B and T-lymphocytes, and the antibodies they generate. Our aim is to formalise some key hypotheses as to the mechanism of resistance. We present a series of simple simulations illustrating changes to the dynamics of the immune response under these hypotheses. These include attenuated viral cell entry, pre-existing cross-reactive humoral (antibody-mediated) immunity, and enhanced T-cell dependent immunity. Finally, we illustrate the potential application of this sort of model by illustrating variational inversion (using simulated data) of this model to illustrate its use in testing hypotheses. In principle, this furnishes a fast and efficient immunological assay – based on sequential serology – that provides a (i) quantitative measure of latent immunological responses and (ii) a Bayes optimal classification of the different kinds of immunological response (c.f., glucose tolerance tests used to test for insulin resistance). This may be especially useful in assessing SARS-CoV-2 vaccines.

## 1 – Introduction

Recently, we produced a series of technical reports applying dynamic causal modelling (DCM) to epidemiological data to illustrate how this might be applied to understanding the Covid-19 pandemic [1-4]. This is an approach developed in neuroimaging research that attempts to find the best explanations for observed timeseries data [5]. An interesting outcome of this work is that the best explanations of data from the current pandemic appeal to there being sizeable portions of the population who either are not susceptible to the virus, or who have such a mild course of illness that they are unlikely to pass it on. The term 'immunological dark matter' has found some media purchase[1] in describing this finding. However, the idea behind this phrase is simple. The implication is that one or more unmeasured variables confer a degree of non-susceptibility or resistance to severe acute respiratory syndrome coronavirus 2 (SARS-CoV-2). This may be partly due to demographic factors, such as geographical isolation or shielding. In addition, it is likely that heterogeneity in the immune response people mount to the virus contributes to this [6-9].

---

[1] https://www.theguardian.com/world/2020/may/31/covid-19-expert-karl-friston-germany-may-have-more-immunological-dark-matter

This raises the question as to what the differences in immunity may be and how we could test this. In this paper, we formalise some of the hypotheses raised as to the immunological factors conferring resistance. We have no pretensions of answering this question – which almost certainly has several different concurrently correct answers – but hope to illustrate how the variational methods used in the epidemiological DCM could be repurposed to test hypotheses about the causes of resistance. This may be particularly salient in the setting of vaccine trials, in which a great deal of immunological data may be collected following a timed exposure to viral antigens – and in which the immunity conferred by a vaccine may transcend purely antibody mediated immunity [10,11], which appears to have a relatively low prevalence [12]. As such, the model presented here is designed to include antibody mediated immunity, but also to allow for alternative (e.g., T-cell mediated) forms of resistance. The use of variational inference procedures in this setting allows for fast and efficient inferences at the individual level, quantification of the uncertainty surrounding these estimates, and hypothesis testing as to the best explanation for individual serology results.

The DCM approach relies upon the notion of a generative or forward model that generates measurable data. The idea here is that, by adjusting the parameters of this model, we can find the parameter configurations that minimise any discrepancy between synthetic and measured data, subject to prior beliefs about the plausibility of those configurations. Practically, this proceeds using variational Bayesian inference [13,14]. We make use of a variational Laplace procedure [15] that maximises a lower bound on the marginal likelihood (or model evidence) of data. The bound is known variously as an evidence lower bound (ELBO) or negative free energy. The benefit of the forward modelling approach is that it appeals directly to the mechanisms that generate data. We are interested in measures of immunity that include the viral load, antibody levels, and T-cell responses (as indicated by interferon-$\gamma$ secretion by CD4 and CD8 positive T-cells). This approach allows us to make inferences about immunity that can arise from interacting and multifactored immune responses.

In the following sections, we first set out a generative model of immune responses. Like many epidemiological models [16] – and existing immunological models [17] – this is a compartment model. This means each component of the immune system exists in one of several alternative states that they may transition between. As in the location, infection, symptom, testing (LIST) DCM [1], our generative model appeals to different factors, corresponding to different arms of the immune system. After introducing this model, we set out formalisations of hypotheses as to the causes of resistance. We then illustrate the fitting of this model to (synthetic) data highlighting how these alternative hypotheses could be tested for a specific resistant individual. Finally, we discuss the potential implications of this approach – including the importance of not overinterpreting negative antibody tests [18].

## 2 – A mean-field model of the immune response

The approach we pursue here is based upon a mean-field approximation that separates an immune response into five different factors or modules [19-21]. These are the antibodies (which subdivide into IgG and IgM types), B-cells, T-cells, and the virus itself. Figure 1 summarises the key aspects of the model dynamics. Each factor has a number of levels, with some probability assigned to each level. The dynamics of the model are formulated in terms of the move of probability mass between these levels. For example, antibody synthesis is the process of moving probability mass from the 'absent' level to the neutralising or non-neutralising antibody states. The dashed arrows between factors in Figure 1 show the direction of influence of factors on one another. We will unpack the model by going through these, starting from the reciprocal influence of the virus on T-cells, followed by T-cell mediation of the B-cell response. We then outline the interaction between B-cells and the virus, B-cells and antibodies, and antibodies and virus.

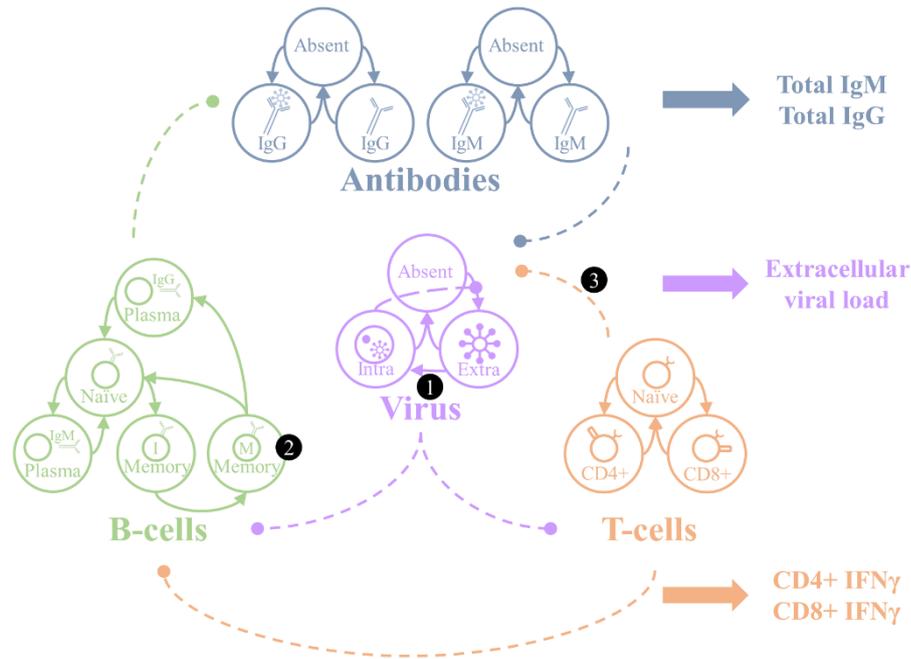

**Figure 1 – The generative model.** This schematic sets out the five factors of the generative model and the data modalities they generate. There are two antibody factors (IgG and IgM). For these factors, the absent antibody can become either neutralising or non-neutralising, and each of these can decay back to being absent. Similarly, the T-cell factor allows for transitions between being naïve to either CD4+ or CD8+ cells. These decay back to naïve, under the assumption that the production of naïve T-cells occurs at approximately the same rate as the death rate of specialised T-cells to maintain a roughly constant T-cell population. The viral dynamics include an absent state, which transitions to being extracellular depending upon the amount of intracellular virus. This represents the intracellular replication and shedding of virus into the extracellular compartment. The extracellular virus then moves into the intracellular compartment – where further replication and shedding can take place. The B-cell factor is slightly more complex as this includes five different levels. Naïve B-cells become either IgM plasma cells or immature memory cells. Immature (I) memory cells develop into mature (M) memory cells, which themselves can become IgG plasma cells. Mature memory cells and plasma cells decay into naïve cells over time. The dashed connections between factors with round arrows express the directed influences between factors, as described in more detail in the main text. The influence of one factor over another is to increase or decrease the rates of transition between selected levels of that factor. The three numbers in black circles indicate the parameters we will manipulate to augment or attenuate the immune response.

The influence of the virus on the T-cells is to increase the rate at which naïve T-cells specialise [22]. Some proportion ($\theta_{CD4}$) of naïve cells transition to being CD4+ cells each hour, scaled by the probability of the virus being extracellular. Similarly, some proportion ($\theta_{CD8}$) of naïve cells transition to being CD8+ cells each hour, scaled by the probability of the virus being intracellular. Our definition of the term 'naïve' T-cells is deliberately broad, to account for any T-cell that has the potential to become an activated CD4+ or CD8+ cell selective for a viral antigen. This might include CD4+ or CD8+ cells that have not become activated or started to secrete cytokines like interferon-γ (INF-γ), in addition to cells that have yet to express CD4 or CD8 surface markers [23].

The influence of the T-cells on the viral factor are as follows. In proportion to the number of CD4+ T-cells, the extracellular virus moves to the absent state. This stands in for the effect of CD4+ cell mediated activation of parts of the innate immune system – for example, the activation of macrophages that phagocytose and kill pathogens [24]. Responses mediated by these cells are known to have a role in influenza virus and coronavirus infections [25,26]. The CD8+ cell causes a move of intracellular virus to the absent state. This represents the cytotoxic activity of these cells, prompting apoptosis of infected cells [27]. The proportion moving to the absent state is given by the probability of a T-cell being a CD8+ cell multiplied by $0.5(\theta_{TCP} + \theta_{dpi})$, where the first parameter deals with the effect of the cytotoxic (CD8+) cell on the pathogen, and the second is the decay of the intracellular pathogen due to T-cell independent mechanisms. The latter is invariant to the T-cell state. We have not specified the mechanism of decay of intracellular and extracellular virus here, such that these variables could encompass innate immunity, non-functional viral mutations, and host cell death [28,29].

In addition to their direct effects on the virus, the T-cells promote a B-cell response [30]. Specifically, in the presence of CD4+ cells, naïve B-cells become (immature) memory cells [31] or IgM-secreting plasma cells. The proportion of B-cells specialising is proportional to the number of CD4+ cells, with proportionality constant $\theta_{BC}$. The proportion of specialising B-cells becoming memory cells is parameterised as $\theta_{BCm}$. The decay of the (IgM and IgG) plasma cells is parameterised as $\theta_{dpB}$ and is independent of T-cells. Similarly, there is a constant move of immature memory cells into the mature memory cell state ($\theta_{mba}$) and a decay of mature memory cells ($\theta_{dmB}$) – neither of which depends upon the T-cells or presence of pathogen. The viral influence on B-cells is to cause the mature memory cells to transition to the IgG plasma cell state (parameterised by $\theta_{Bmp}$) [32].

The B-cell factor exerts an influence over the antibody factors, with IgM plasma cells leading to a move from absent to neutralising or non-neutralising proportional to $\theta_{pAb}$ (where the proportion that are neutralising is given by $\theta_{nAb}$) in the IgM antibody factor, and IgG plasma cells leading to the equivalent dynamics in the IgG antibody factor. Regardless of the B-cell factor, antibodies decay to the absent state with rate parameterised by $\theta_{dAb}$ for IgM antibodies and the slower $\theta_{dAbG}$ for IgG antibodies. Finally, the antibodies exert an influence over the viral population, with extracellular virus moving to the absent state in proportion to the antibodies in the neutralising state. The overall structure of these dynamics is given in Equation 1, where the $\Delta\tau$ used for the simulations is one hour.

$$\begin{aligned}
\mathbf{p}(\tau+\Delta\tau)^i &= \mathbf{T}(\theta)^i \left( \mathbf{p}(\tau)^V \otimes \mathbf{p}(\tau)^{TC} \otimes \mathbf{p}(\tau)^{BC} \otimes \mathbf{p}(\tau)^{IgM} \otimes \mathbf{p}(\tau)^{IgG} \right) \\
&i \in \{V, TC, BC, IgM, IgG\} \\
\mathbf{p}(0)^V &= \begin{bmatrix} 1-e^{-8}\theta_m & e^{-8}\theta_m & 0 \end{bmatrix}^T \\
\mathbf{p}(0)^{TC} &= \begin{bmatrix} 1 & 0 & 0 \end{bmatrix}^T \\
\mathbf{p}(0)^{BC} &= \begin{bmatrix} 1-e^{-4}\theta_n & 0 & 0 & e^{-4}\theta_n & 0 \end{bmatrix}^T \\
\mathbf{p}(0)^{IgM} &= \begin{bmatrix} 1 & 0 & 0 \end{bmatrix}^T \\
\mathbf{p}(0)^{IgG} &= \begin{bmatrix} 1 & 0 & 0 \end{bmatrix}^T
\end{aligned} \quad (1)$$

Each **p** is a vector of probabilities, with superscripts indexing the factor that these probabilities pertain to. **T** is a transition matrix that updates these probabilities for a given time interval. The order of states shown in these probability vectors are as follows. The viral states are *absent*, *extracellular*, and *intracellular*. The T-cell states are *naïve*, *CD4+*, and *CD8+*. The B-cell states are *naïve, plasma (IgM), immature memory, mature memory, plasma (IgG)*. Both sets of antibody states are *absent, non-neutralising, neutralising*. Note that a very small amount of virus can be present in the extracellular space at the start, and that a small proportion of B-cells exist in the mature memory B-cell state at the start.

For a generative model to be useful, it should be able to generate data of the sort that can be measured. The model in Figure 1 generates antibody levels, extracellular viral load, and the proportion of activated CD4+ [33,34] and CD8+ cells [7]. These are generated by summing the proportions of neutralising and non-neutralising antibodies, and taking the proportions of the other variables, and then scaling these by the maximum values we expect them to take. The scaling factors include a fixed and a random component. The fixed component determines the maximum value when the random component takes its prior expected value of 1. The random variables are $\theta$-parameters as detailed in Table 1, meaning these could be estimated from measured data. These measurements are relative to some baseline which is likely to be zero for viral load but may be higher for the other variables. IFN-γ secretion is generally accepted as a measure of activation of CD4+ and CD8+ cells [35]. Evidence from H7N9 influenza virus infections suggests that recovery from the viral infection is positively associated with IFN-γ secretion by CD4+ and CD8+ cells [35]; emphasising the importance of these 'activated' cells as measures of the immune response. Put formally, the process generating data is as follows:

$$\begin{aligned}
\mathbf{y}(\tau)^{IgM} &= 2^7 \left( \mathbf{p}(\tau)_2^{IgM} + \mathbf{p}(\tau)_3^{IgM} \right) \theta_{Abm} \\
\mathbf{y}(\tau)^{IgG} &= 2^7 \left( \mathbf{p}(\tau)_2^{IgG} + \mathbf{p}(\tau)_3^{IgG} \right) \theta_{Abm} \\
\mathbf{y}(\tau)^{VL} &= 10^4 \mathbf{p}(\tau)_2^{V} \theta_{VLm} \\
\mathbf{y}(\tau)^{CD4} &= \mathbf{p}(\tau)_2^{TC} \theta_{IFN} \\
\mathbf{y}(\tau)^{CD8} &= \mathbf{p}(\tau)_3^{TC} \theta_{IFN}
\end{aligned} \quad (2)$$

Each of these data modalities (**y**) are multiplied by a constant and a scaling parameter, which may be estimated during model inversion. By inversion, we mean the process of inferring the parameters that caused the data. We do not assume data are generated every hour (the rate of update of latent dynamics). Instead, we assume measurements are only available every 24 hours. This is perhaps a little optimistic, but the model is formulated such that the measurement timings can be arbitrarily set, to account for when measurements are available. Table 1 provides the parameters used in the simulations that follow. It is important to emphasise that the priors chosen here are largely heuristic. These will vary with infection between different organisms, and between hosts. This means the simulations that follow should not be taken as making any comment on any specific infection. Ideally these would be estimated from empirical data. In the absence of these data, we have selected parameters such that the relative timing of each response is broadly consistent with that observed in biology, and that is consistent with a self-limiting infection that is largely cleared within about three weeks. We discuss this consistency in relation to the time course of measurements during SARS-CoV-2 infections below. While this allows us to illustrate general principles, these illustrations should be taken as qualitative as opposed to quantitative.

**Table 1 – Priors for normal immune response (log parameters)**

| Parameter | Symbol | Expectation | Variance |
|---|---|---|---|
| Scaling for initial immunity | $\ln \theta_n$ | -2 | 1 |
| Scaling for initial viral load | $\ln \theta_m$ | 0 | 1/256 |

| Decay of IgM antibodies | $\ln \theta_{dAb}$ | -6 | 1/1024 |
|---|---|---|---|
| Decay of IgG antibodies | $\ln \theta_{dAbG}$ | -8 | 1/1024 |
| Production of antibodies by plasma cells | $\ln \theta_{pAb}$ | 0 | 1/1024 |
| Proportion of antibodies that are neutralising | $\ln \theta_{nAb}$ | -1 | 1/1024 |
| Decay of plasma B-cells | $\ln \theta_{dpB}$ | -2 | 1/1024 |
| Decay of memory B-cells | $\ln \theta_{dmB}$ | -32 | 1/1024 |
| Activation of memory B-cells | $\ln \theta_{mba}$ | -4 | 1/1024 |
| Proportion of mature memory cells becoming plasma cells in presence of infection | $\ln \theta_{Bmp}$ | -1/2 | 1/1024 |
| Specialisation of naïve B-cells in the presence of CD4+ T-cells | $\ln \theta_{BC}$ | -3 | 1/1024 |
| Proportion of B-cells specialising as memory cells | $\ln \theta_{BCm}$ | -1 | 1/1024 |
| Decay of CD4+ cells | $\ln \theta_{dT4}$ | -4 | 1/1024 |
| Decay of CD8+ cells | $\ln \theta_{dT8}$ | -5 | 1/1024 |
| Production of CD4+ cells in presence of extracellular pathogen | $\ln \theta_{CD4}$ | -3 | 1/1024 |
| Production of CD8+ cells in presence of intracellular pathogen | $\ln \theta_{CD8}$ | -3 | 1/1024 |
| Neutralisation of intracellular pathogen by T-cell mediated apoptosis | $\ln \theta_{TCP}$ | -2 | 7/32 |
| Neutralisation of extracellular pathogen by antibody independent CD4+ mechanisms | $\ln \theta_{T4P}$ | -8 | 1/1024 |
| Viral entry into cells | $\ln \theta_{int}$ | -2 | 1/16 |
| Viral shedding into extracellular space (absorbing replication rate) | $\ln \theta_{ext}$ | -1 | 1/1024 |
| Decay of extracellular pathogen (independent of adaptive immunity) | $\ln \theta_{dpe}$ | -2 | 1/1024 |
| Decay of intracellular pathogen (independent of adaptive immunity) | $\ln \theta_{dpi}$ | -2 | 1/1024 |
| Scaling of antibody titre measures | $\ln \theta_{Abm}$ | 0 | 1/256 |
| Scaling of viral load measures | $\ln \theta_{VLm}$ | 0 | 1/256 |
| Scaling of IFN-γ T-cell assays | $\ln \theta_{IFN}$ | 0 | 1/256 |

While the variances in Table 1 may seem very small, it is important to remember that we are working with log parameters – so small changes can result in large changes once exponentiated. We will see the large confidence intervals in subsequent figures reflecting this. Secondly, many of the parameters we deal with play the role of probabilities, which means tight priors are required so that these parameters do not exceed a sum total of one[2]. We use these tighter priors for simulation purposes; we relax the variances later, when demonstrating model fitting. It is worth drawing attention to the variances for the initial immunity, the viral cell entry, and the T-cell mediated apoptosis parameters. The variances for these are chosen to be very wide as these are the key parameters we will be varying later. Specifically, these are chosen such that the modifications we will make to these parameters amount to a change of approximately one standard deviation each[3].

---

[2] This highlights a trade-off in our decision to use normally distributed priors. An alternative would be to use Dirichlet distributed priors, which ensure summation to one. However, the flexibility afforded by the use of normal distributions, and the simplicity of variational inference under a Laplace assumption, favours the use of normally distributed log parameters.

Figure 2 illustrates an immune response generated from the prior expectations in Table 1. Here we see the rapid increase in viral load as the initial extracellular virus enters the cells enabling intracellular replication and shedding into the extracellular space. As the extracellular viral load increases, this initiates a helper (CD4+) T-cell response that increases the number of plasma B-cells and causes a rapid increase in antibody levels. Later peaks are seen in the populations of cytotoxic (CD8+) T-cells and memory B-cells. As the viral load is suppressed, the immune cell populations decay, with the exception of the mature memory cells. This is because, in the presence of the virus, mature memory cells rapidly transition into plasma cells. This means they are the only cell type to decrease in number during the initial phase of the infection but increase later on from the immature memory cells acquired during the infection. In the antibody response plot, note the early IgM response, followed by a slower but more sustained IgG response. This reflects the characteristic role of these isotypes in fast and long-lasting immunity, respectively.

Figure 2 also illustrates some face validity of this approach in relation to the emerging data on the immune response in Covid-19. The time-course and magnitude of the antibody response is broadly consistent with data suggesting peak antibody titres in the region of about $2^6$ [36] (measured as chemiluminescence levels relative to some cut-off), consistent with the upper bound approaching 60 in the antibody plot in Figure 2, and an increase in IgG relative to IgM over time, both plateauing at around day 20 following symptom onset. This is slightly earlier than in our simulation but may reflect the time it takes for the viral load to reach a level that causes symptoms. For ease of comparison with empirical reports we plot the viral load ($\mathbf{y}^{VL}$), generated according to Equation 2, after transforming into the cycle threshold for detection using a polymerase chain reaction (PCR). See the legend of Figure 2 for details. The viral load appears to decline from a cycle threshold (Ct) of about 30 to 40 over the space of about 10 days [37], consistent with the lower left plot in Figure 2. In the lower right plot, the percentage of CD4+ and CD8+ cells producing IFN-$\gamma$ peaks at around 0.2 per hundred cells (with upper limit around 0.5), consistent with the kinds of ranges reported in some empirical studies [38]. While these comparisons are heuristic and will almost certainly vary with laboratory equipment and local measurement procedures, they illustrate face validity in the sense that the data generated broadly resemble measurable data.

---

[3] This accounts for the variances of 1, 7/32, and 1/16, which are much larger than those for other priors. The reason for this will become more obvious in Section 4. Briefly, they are chosen to avoid biasing our model fitting towards one phenotype or another.

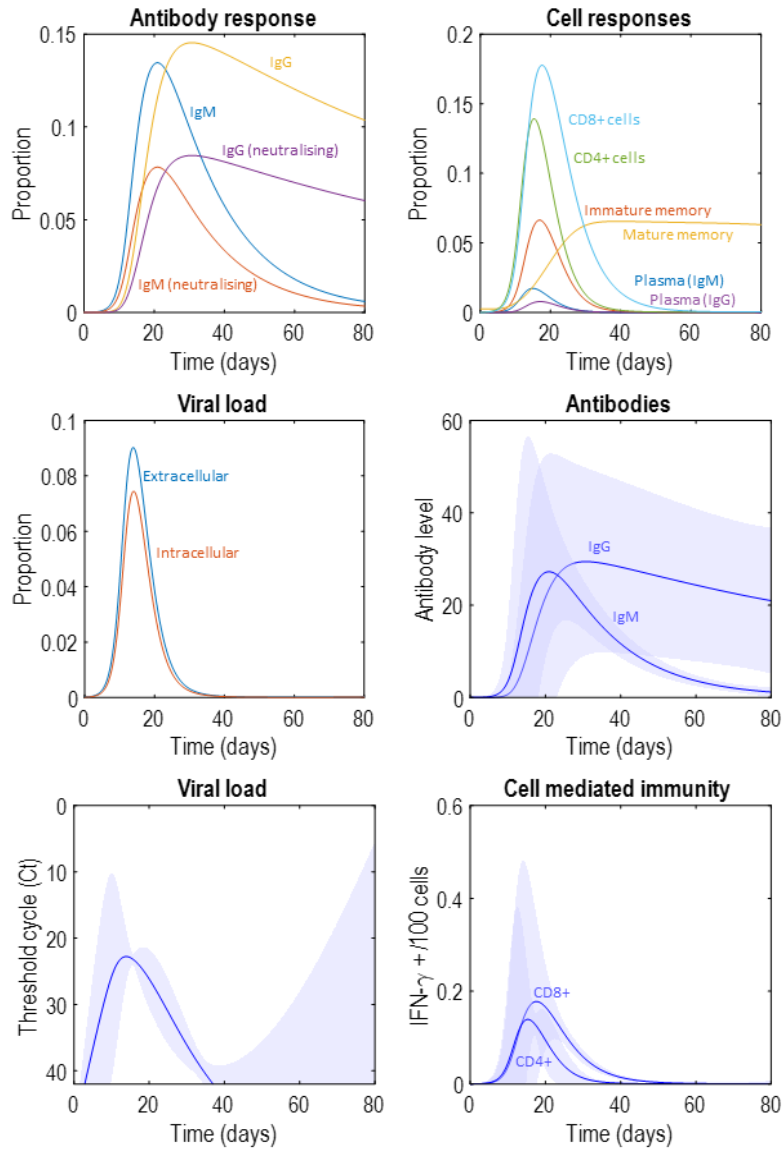

**Figure 2 – A synthetic immune response.** The *upper left* plot shows the latent variables in terms of the different types of antibody. This distinguishes between the faster IgM response, and the more prolonged IgG response, and between the neutralising and non-neutralising variants of each. The *upper right* plot shows the cellular latent variables, with an initial increase in the activated T-cell proportions that leads to a later increase in plasma and memory cells, with a persistent memory response developing. The *middle left* plot shows the viral states, subdivided into intracellular and extracellular components. The *middle right* plot shows the total antibody level, including both neutralising and non-neutralising. The *lower left* plot shows the viral load in terms of the 'threshold cycle' (Ct) which indicates the number of cycles of polymerase chain reaction (PCR) required before the viral nucleic acids are detectable. A greater number of cycles indicates a greater dilution (i.e., a smaller concentration). To generate these data, we took the negative log [39,40] of $\mathbf{y}(\tau)^{VL}$, scaled it by a factor of 4, and added a constant 50 – under the assumption that more than 40 cycles implies a negligible viral concentration, based on fluorescence amplification results for a typical qPCR run [41]. These scale factors and constants could in principle be fit to empirical data. Here they are used only for the plotting and have no influence on model inversion. Finally, the *lower right* plot shows the proportion of CD4+ and CD8+ cells releasing IFN-γ.

### 3 – Routes to resistance

Figure 3 provides an overview of the mechanisms for resistance that we address here. This is not exhaustive but highlights three important ways in which the immune system might vary between individuals. We will take each of these in turn and demonstrate the influence they have on the dynamics of the immune response under our model. Figure 4 offers a comparison between the dynamics of latent states under each manipulation for ease of comparison between the time course of antibody, cellular, and viral responses. The same plots will be shown in subsequent figures, which we will unpack in detail.

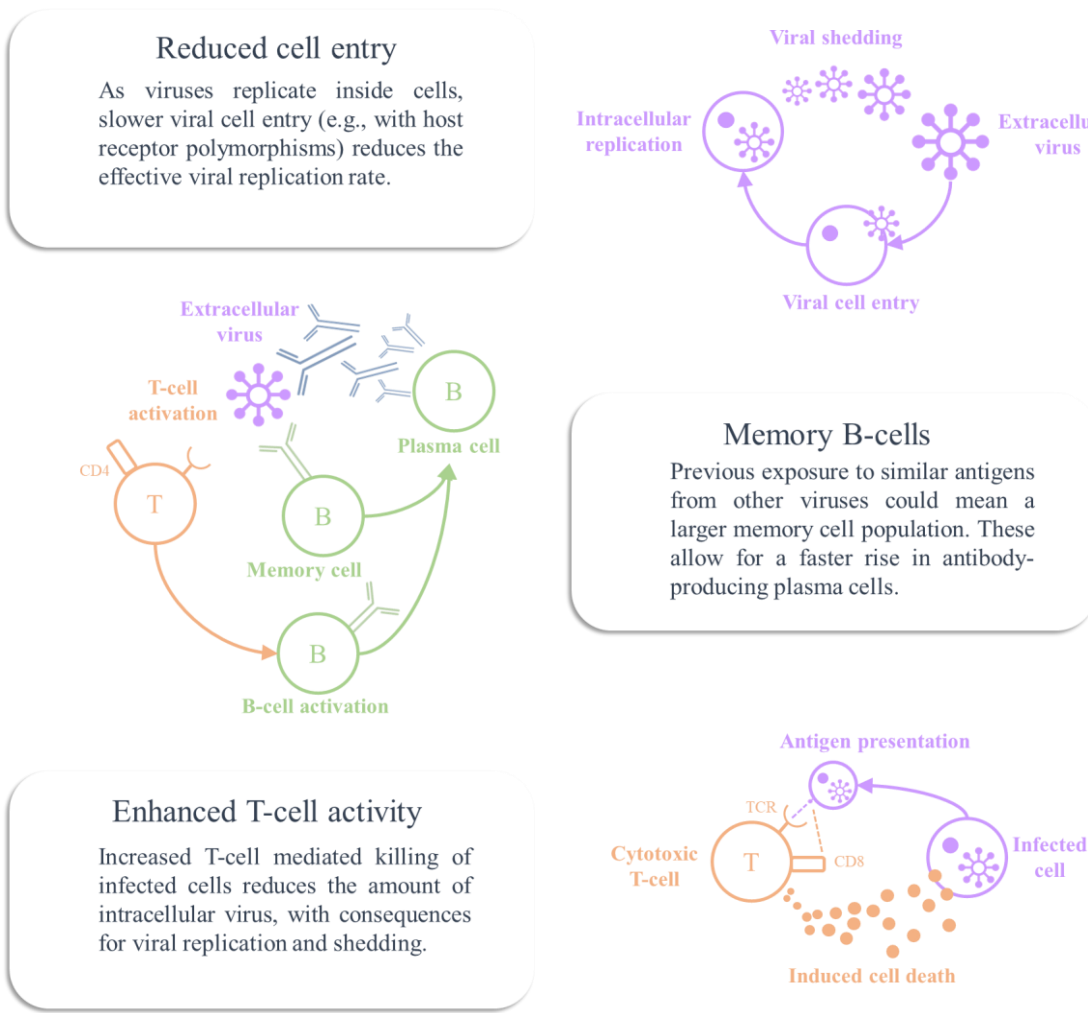

**Figure 3 – Mechanisms of resistance.** In what follows, we will illustrate various modifications to the model simulated in Figure 2, by changing model priors. The graphics in this figure provide an overview of the three manipulations we will pursue in terms of the associated biological mechanisms. These mechanisms are reduced viral entry into host cells, pre-existing memory B-cells, and enhanced T-cell dependent killing of infected cells.

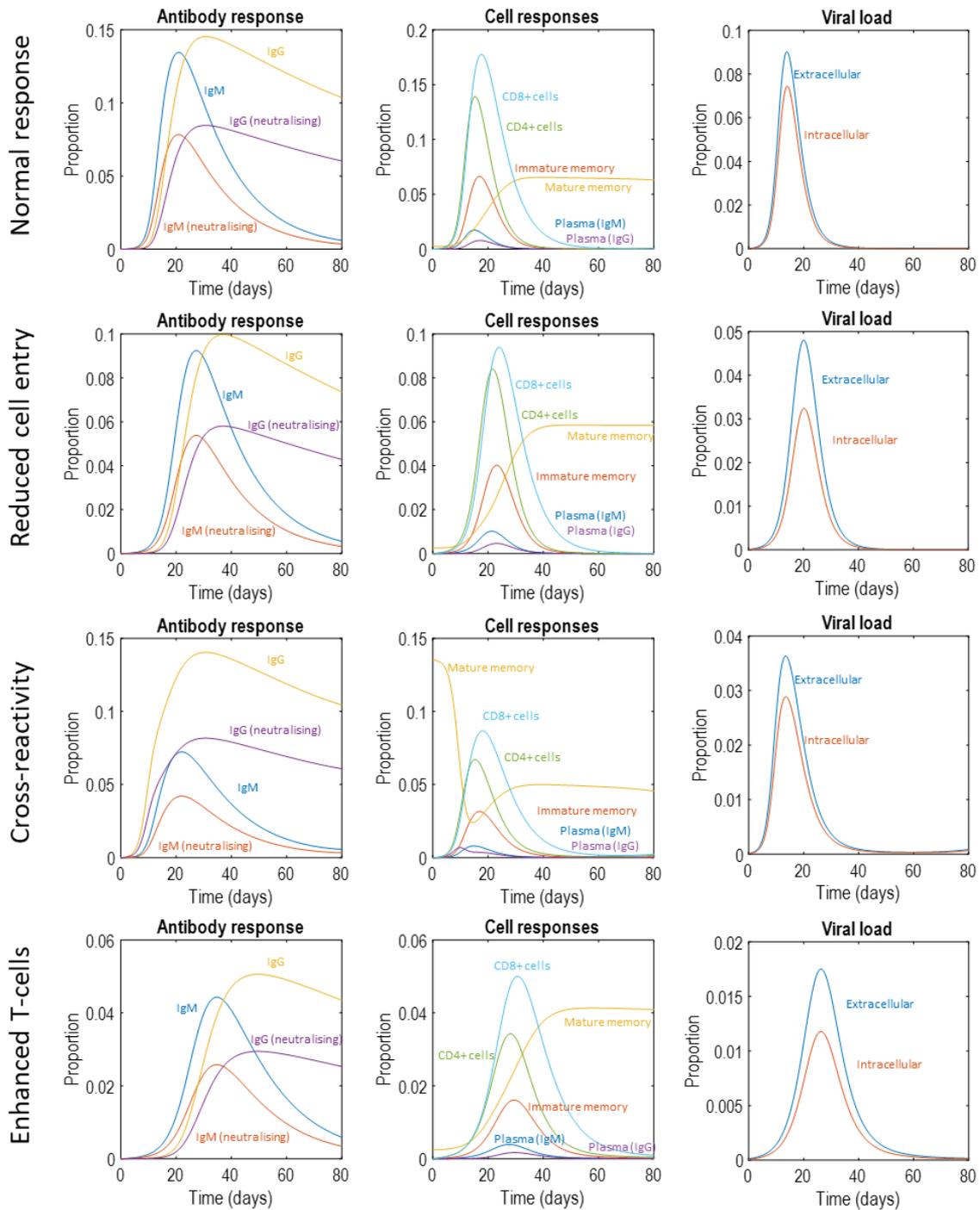

**Figure 4 – Latent variables.** This figure takes the latent variable plots from Figure 2 and supplements them with those that will appear in Figures 5-7. The format is as described in Figure 2. We will unpack the details of these plots as they are introduced in subsequent figures. However, the highlights are (i) the attenuation of peak viral load in the second to fourth row of plots, relative to the first row, (ii) the smaller antibody response required to suppress the infection in the second and fourth rows of plots, (iii) the earlier peak in IgG plasma cells and antibodies in the third row of plots. The key message from this figure is that there are several different dimensions along which immune responses may vary, with consequences for viral suppression.

The first route to resistance is a reduced capacity for viral entry into cells. In the context of SARS-CoV-2, this could be due to polymorphisms in the cell surface ACE-receptors targeted by the virus [42] – or to age-dependent differences in expression [43]. In terms of the model outlined above, this is an attenuation of the rate at which extracellular virus becomes intracellular. Specifically, we changed the expectation of $\ln\theta_{int}$ from -2 to -2.2. Figure 5 shows a simulation of this scenario. Given that viral reproduction occurs in the intracellular compartment, slower entry into this compartment leads to a failure to increase the viral population. This leads to a viral load that is lower at its peak than in Figure 2 – possibly suggesting people with this phenotype are less infectious (note the differences in y-axis scales between Figures 2 and 5). However, the most obvious difference between Figures 2 and 5 is the magnitude of the immune response required to clear the virus. In Figure 5, we see a much smaller response, and might anticipate a failure to detect antibodies post-infection in such individuals. In addition, a smaller immune response might mean fewer symptoms during the course of the infection.

The second route is pre-existing cross-reactivity of antibodies, which we simulate through assuming a larger population of memory cells capable of rapid specialisation into plasma cells that target the virus. By cross-reactivity, we mean that antibodies developed to some previous infection may also be effective against the current infection [8,44]. Specifically, we formalise this by increasing $\ln\theta_n$ from -2 to 2. Figure 6 illustrates the rapid drop in mature memory cell populations as they become IgG plasma cells, and the resulting increase in IgG antibodies. Note that the plasma IgG cells never reach the same level as the original memory cell population. This is because their rate of decay is considerably faster than that of memory cells. The key differences between this simulation and that reported in Figure 2 are the earlier and more dramatic IgG response, which now precedes the attenuated IgM response, and the decrease in peak viral load. Longitudinal studies of these antibodies in Covid-19 show this pattern in patients with low severity disease, compared to an earlier IgM response (as in Figure 2) in more severe disease [45].

Third, the response could be mediated by antibody independent T-cell mechanisms [7,46] that suppress the virus before the viral load is sufficient to promote plasma cell proliferation. A similar hypothesis could be advanced for a purely innate immune response, by increasing the rate of decay of the virus independently of the states of the other three factors. Cell-mediated immunity is well recognised to vary with age [47,48] and could account for the relative resistance of younger members of the population. Suggestive evidence in favour of the relevance of CD8+ dependent mechanisms comes from genomic studies that have identified loci associated with lung CD8+ cells as differing between those with severe Covid-19 and the control population (people without a known diagnosis of Covid-19) [49]. Some T-cells in the respiratory system are thought to have a role in immunological memory, which may be independent of B-cell memory systems [50,51]. This is one possible interpretation for a more vigorous T-cell response in some people, whose T-cells have encountered similar antigens in the past. There are also emerging data suggesting SARS-CoV-2 itself induces a memory T-cell response [38], in addition to there being T-cells reactive to viral antigens in people who have yet to be exposed [7]. Although we have formalised this in terms of CD8+ responses (by increasing $\ln\theta_{TCP}$ from -2 to -1/8), there is also evidence in favour of a role for CD4+ responses in rapid suppression of influenza viruses in some individuals – something that has been proposed as a mechanism for the difference in disease severity between successive infections in individuals [26]. Figure 7 shows the influence of increasing the effectiveness of T-cell dependent killing on the immune response, which appears to reduce viral load and dampen and prolong antibody and cell mediated responses.

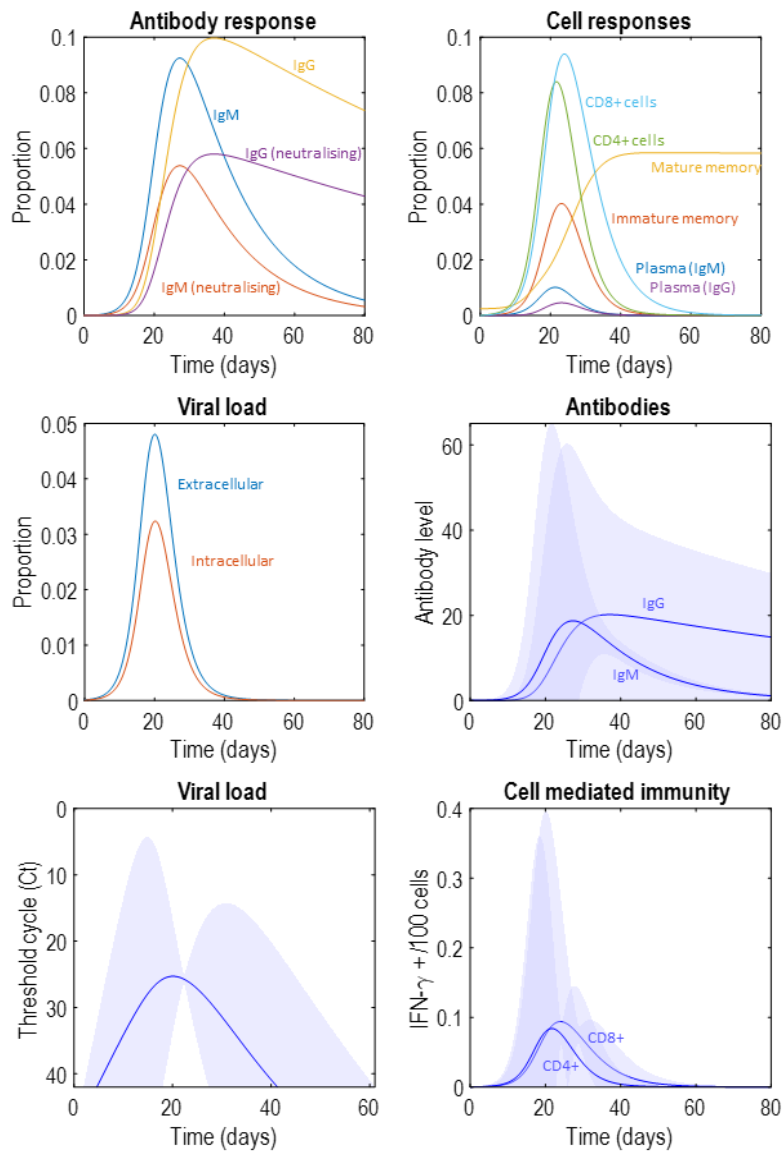

**Figure 5 – Attenuating viral entry into cells.** This figure uses the same layout as Figure 2. Here, the proportion of extracellular virus entering the intracellular compartment per hour has been decreased, remembering that the production of extracellular virus depends upon there being intracellular virus that can use the cellular machinery to manufacture and secrete new virus particles. Decreasing the rate of cell entry therefore slows the growth of the viral population, leading to a smaller peak with a smaller immune response required to suppress the infection. Note the decrease in antibody titre, which might mean that people exhibiting this immune phenotype may not have detectable antibodies in serological testing post-infection.

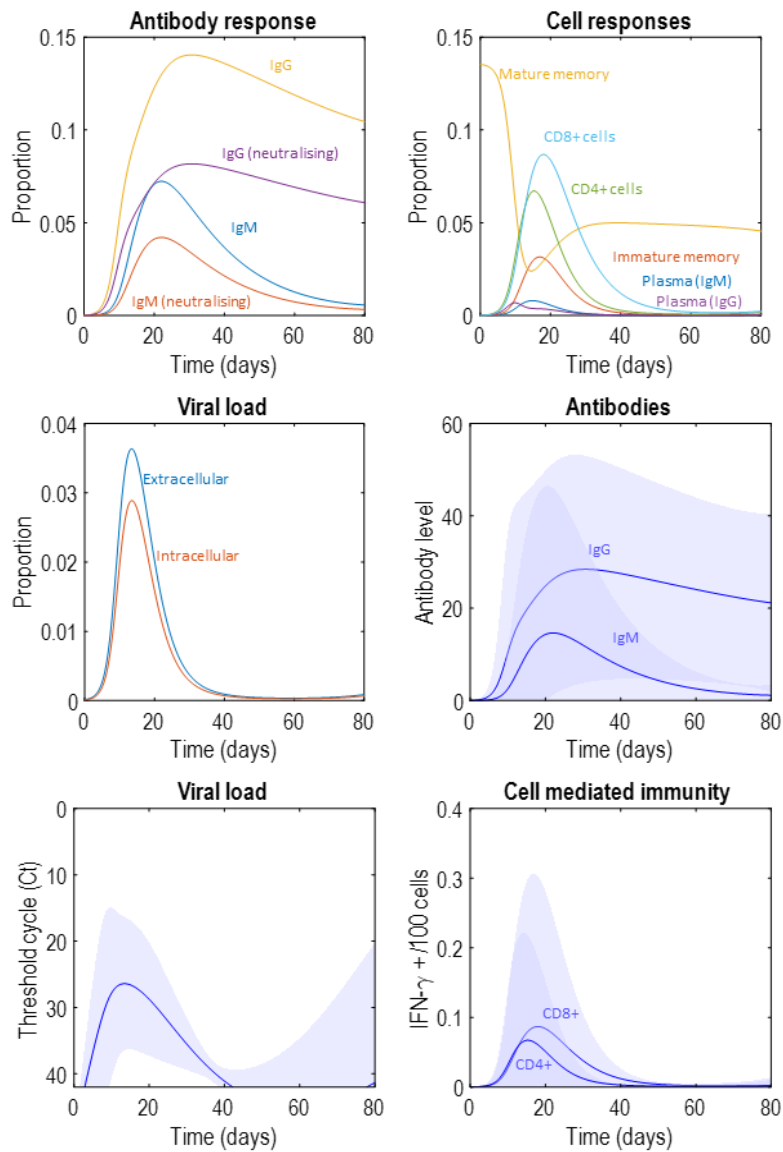

**Figure 6 – Pre-existing immunity.** This figure uses the same layout as Figure 2. Here, the proportion of mature memory cells at the start of the infection has been increased, to simulate the scenario that immunological memory may have been instantiated by exposure to related viruses in the past. Note the rapid increase in the IgG response, in contrast to the early IgM response we saw in Figure 2. Like in Figure 5, the virus is suppressed and never reaches as high a viral load as in Figure 2. However, here we see a sustained increase in serum immunoglobulin in the latent (upper left) and measurable (middle right) antibody response plots. This pattern of response might be associated with a very mild illness, not prompting testing for the virus (or possibly even an undetectable viral load on being tested) but would be associated with positive serological tests later on.

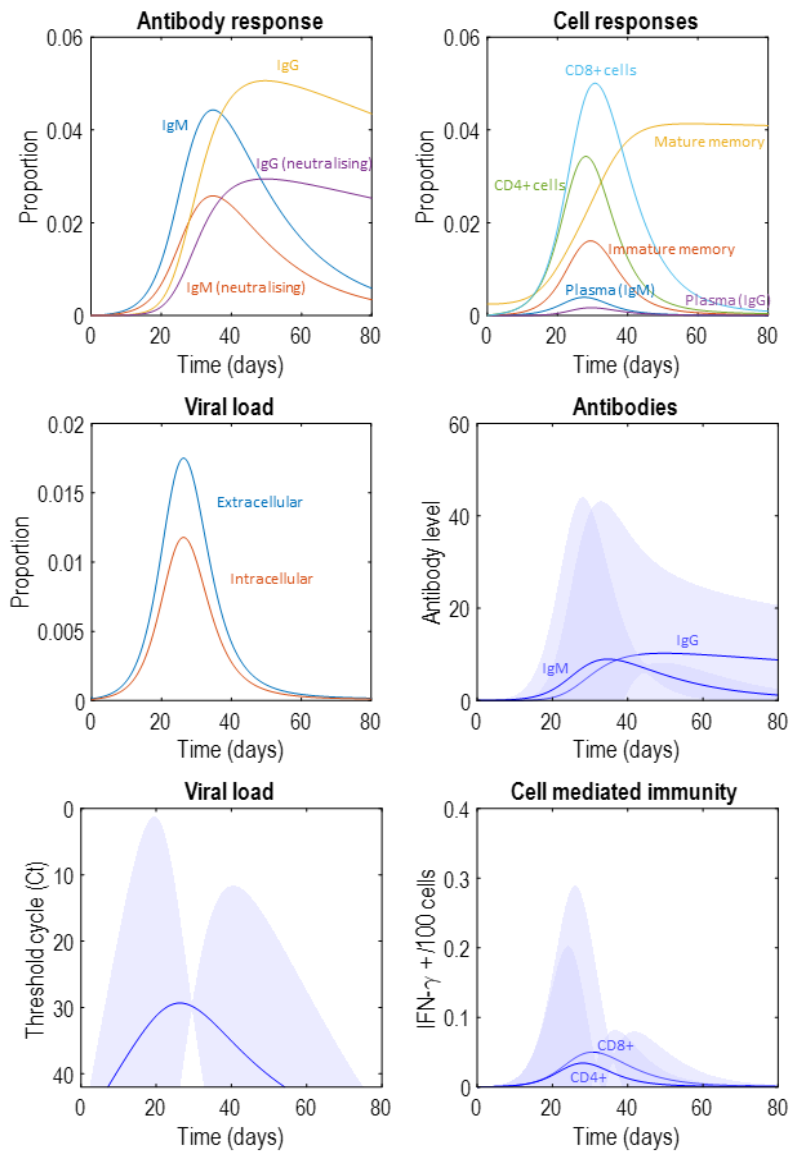

**Figure 7 – Cell mediated immunity.** This figure uses the same layout as Figure 2. The scenario simulated here is one in which the proportion of intracellular virus destroyed through T-cell mediated mechanisms is increased. This could be interpreted as a memory T-cell mediated response. Here we see a smaller viral load and a flattened antibody response.

Clearly there are many other phenotypes we could have considered [52]. These could include mixtures of the above, or more extreme values of the parameters. For example, if we reduce the viral cell entry sufficiently, the viral load never rises at all, but immediately decays to zero – with a negligible immune response. In contrast, if we sufficiently attenuate the immune response (e.g., by reducing activation of T and B cells), the viral load increases to its carrying capacity and stays there. There are a range of other intermediate scenarios, like a small but persistent viral load that is never fully suppressed (perhaps more consistent with some hepatitis viruses [53]), or even second waves (without reinfection) if the antibodies decay sufficiently quickly. These parameter settings might be more appropriate in modelling viruses with latent phases that reactivate when immunity decays. For example, varicella zoster virus causes chickenpox before being suppressed, but can re-emerge later as shingles [54].

## 4 – Immunological phenotyping

In this section, we attempt to do two things. The first is to find out whether the three phenotypes identified above (lower proportion of extracellular virus entering cells; higher proportion of mature memory cells on infection; and higher proportion of intracellular virus destroyed through T-cell mediation) may be disambiguated from one another based upon the data that we assume they generate. This rests on the idea that an individual has one, rather than a mixture, of these phenotypes. However, it would be simple to extend this analysis by specifying additional phenotypes that comprise mixtures of those outlined in Section 3. The second aim in this section is to ask whether we could identify interactions between demographic factors – such as whether or not someone has received a Bacillus Calmette-Guérin (BCG) vaccine – and parameters of the models above – for example, T-cell dependent killing of pathogen [55]. We will use the BCG hypothesis, that T-cell dependent killing is enhanced by the BCG vaccine, as an example to aid intuition. This is not intended to endorse or refute this hypothesis, but to show how DCM may be used to formulate and test hypotheses of this sort.

For the first of these aims, we took the outputs of the above simulations assuming one measurement per day for the first three weeks, added univariate random fluctuations with a variance of 1/8 to the square roots of the data. We then fit each synthetic dataset – comprising daily measurements of the viral load, antibody levels, and proportion T-cells releasing of IFN-γ – to the model with the prior expectations in Table 1, but with variances multiplied by 16. Model fitting here means finding approximate posterior probabilities $Q$ that maximise a lower bound on model evidence (the marginal likelihood of the data given the model):

$$Q(\theta | m) = \underset{Q(\theta|m)}{\arg\max} \left\{ \mathbb{E}_{Q(\theta|m)} \left[ \ln P(\theta, \mathbf{y} | m) - \ln Q(\theta | m) \right] \right\} \approx P(\theta | \mathbf{y}, m)$$

$$\mathbb{E}_{Q(\theta|m)} \left[ \ln P(\theta, \mathbf{y} | m) - \ln Q(\theta | m) \right] \leq \ln P(\mathbf{y} | m)$$

(3)

In Equation 3, $\theta$ are the parameters, $\mathbf{y}$ the data, and $m$ is the model. Maximisation of the model evidence ensures an optimal trade-off between the accuracy with which data can be explained and the simplicity of that explanation – penalising excessively complex explanations. This maximisation was achieved using a standard software routine (**spm_nlsi_GN.m**[4]). We then used Bayesian model reduction (BMR) – a method for efficient post-hoc model comparison [56,57] – to evaluate the relative evidence for alternative models with the prior expectations set out in Table 1, with expectations modified as described in Section 3. The reason for the more uncertain priors used during fitting of the full model is that BMR treats each alternative model as a reduced version of the 'full' model. By setting the variances in the full model to be much larger than those in any of the reduced models, we ensured that the expectations of the reduced priors would be plausible under the full prior distribution. The principle that underwrites BMR is that, when only the priors differ between models, the relative evidence of two models is given by the following equation – which only requires inversion of one of the two:

---

[4] Freely available as part of the SPM12 package at https://www.fil.ion.ucl.ac.uk/spm/software/spm12/

$$\ln P(\mathbf{y}\mid\tilde{m}) - \ln P(\mathbf{y}\mid m) = \mathbb{E}_{P(\theta\mid m,\mathbf{y})}\left[\ln\frac{P(\theta\mid\tilde{m})}{P(\theta\mid m)}\right] \qquad (4)$$

The graphic on the left of Figure 8 shows a confusion matrix, whose rows correspond to the model used to fit the data, and whose columns correspond to the model that generated those data. Each element of the matrix is shaded by the posterior probability of the model given the data, with white indicating a probability of one and black zero. Posterior probabilities are obtained through using a softmax (normalised exponential) operator on the log model evidence given by BMR for each column. The lighter shading along the diagonal elements compared to the off-diagonal elements in the same column implies it is possible to recover the mechanism generating observed timeseries – albeit with varying degrees of confidence. The implication of this is that, in principle, it is possible to use empirical data to assign individuals to the immunological phenotype that best explains their measured serology.

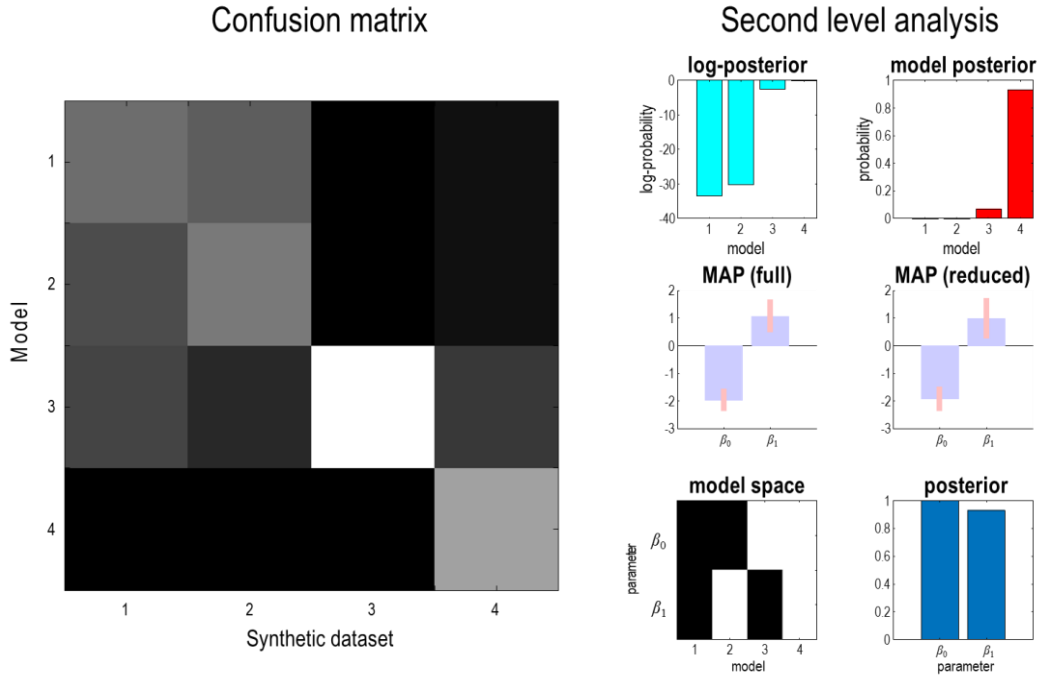

**Figure 8 – Immunological phenotyping.** This figure illustrates two ways in which this modelling approach may be used. The left panel shows a confusion matrix that plots the posterior probability of each model given each dataset. White indicates a probability of one, while black indicates a probability of zero, with grey being intermediate values. The diagonal elements of this matrix represent the posterior probability of a model given the data that it was used to generate, and these probabilities are reassuringly higher than those for other models. The models are ordered such that 1-4 correspond to The rows of Figure 4. The idea here is that alternative hypotheses may be posed to data from an individual's immune response, and that we hope to be able to recover which model best represents that individual's immune response. The second use of this approach is shown on the right, where a synthetic group level analysis has been used to test hypotheses about the role of demographic factors (e.g., whether or not someone has received a BCG vaccine) on parameters of the model (e.g., the cytotoxic T-cell response). The plots labelled *log-posterior* and *model posterior* show the probability associated with group level models that include various combinations of constant ($\beta_0$) and linear ($\beta_1$) effects on the $\theta_{TCP}$ parameter. The constant effect is the expected value for all those without a BCG vaccine, while the linear effect is the amount added to the log expected parameter when someone has had this vaccine. The presence or absence of these parameters (where absence means set to zero) for each model is shown in the *model space* plot with white indicating presence and black absence. The *posterior* plot shows the probability of each parameter being present. The *MAP* plot shows maximum a posteriori estimates for the model including all parameters (with 90% credible intervals) while the *MAP (reduced)* plot shows these when averaged under the relative probability of each reduced model.

The plots shown on the right of Figure 8 illustrate an alternative application of this approach to answer a question at the level of between subject effects. Here, we generated 32 immune responses, as if from 32 different subjects. Each subject was assigned to the BCG group or the unvaccinated group with a probability of ½. The priors were all as in Table 1, except for the $\theta_{TCP}$ parameter whose expectation was sampled from a normal distribution for each individual. The mean of this distribution was set at -2, with 2 added for those with a BCG. In this synthetic example, this implies the BCG vaccine confers some T-cell mediated immunity. We then fit each of the 32 synthetic datasets using the priors of Table 1, and used a parametric empirical Bayes scheme (**spm_dcm_peb.m**) to estimate the parameters of a second level linear model [56,58], which predicted the $\theta_{TCP}$ parameter based upon the presence or absence of the BCG. This second level model has the form:

$$\ln \theta_{TCP}^i = \beta_0 + X^i \beta_1 + \omega^i \qquad (5)$$

Here, the superscript *i* indicates the subject, the $\beta_0$ is a constant value for those with no BCG, and the $\beta_1$ is the effect of having had the BCG. $X^i$ is one if a subject has had the BCG, and zero otherwise. The $\omega$ term indicates normally distributed, zero-mean, fluctuations. The plots on the right of Figure 8 show that a model in which both $\beta$ parameters are allowed to be non-zero provides the best explanation for the synthetic data. Furthermore, the maximum *a posteriori* estimates are consistent with the -2 for the absence of a BCG and the positive effect associated with this vaccine used to generate the synthetic data. This illustrates that, if it were true that the BCG vaccine enhances T-cell-mediated immunity, we could recover this information using this sort of analysis. Again, this is not to suggest that this hypothesis is correct (or incorrect). Instead, our aim is to show that it could be demonstrated if it were true.

## 5 – Discussion

How does this sort of modelling relate to epidemiology? Drawing from theoretical neurobiology, we could think of immunological and epidemiological models as simply different levels of parcellation – just as we have the option of parcellating the brain on the level of regions or on the level of networks, depending upon the question at hand. Recent work [59] in this direction makes use of renormalisation groups [60] to translate between these scales. In brief, this works by identifying functional units at small spatial scales (e.g., individuals at different stages of viral exposure) and extracting low dimensional summaries of their dynamics based upon the system's Jacobian. These are typically along the direction of the eigenvectors with the smallest negative eigenvalues – effectively omitting dynamics that dissipate very quickly [61]. This procedure may be applied recursively to move to high level summaries in a population (e.g., the proportion of individuals at each stage). A key feature of these analyses is that increases in spatial scales are accompanied by slower dynamics. Consistent with this, the course of infection for an individual – lasting for a period of a few days to weeks – is much shorter than that of an epidemic (a few months) or a pandemic (potentially years). This method affords the possibility of using inferences drawn at the level of individual immune responses to nuance models used at an epidemiological level.

It is important to emphasise that none of the hypotheses considered above are 'new'. All have been advanced in the context of the current pandemic or in other infections. What is offered here is a formalisation of these hypotheses, and a generative model that could be used to evaluate them. In addition, generative models may be used to simulate interventions (e.g., pharmacological treatments) to try to understand what might happen before moving to more expensive or risky empirical tests. The generative modelling approach has been employed in the past to assess models of the immune response to dengue virus [62]. The authors used temporally dense measurements of viral load and antibody titre to fit a 'target-cell' model – highlighting the utility of dynamic modelling of immunity. The approach pursued in this paper is closely related to the target-cell approach to modelling viral infections but differs in a few important ways [63]. Target-cell models – in their simplest manifestation – describe the dynamics of an infection in much the same way Susceptible-Exposed-Infectious-Recovered (SEIR) models describe an epidemic. In place of the number of individuals at various stages of being infected by a pathogen, target-cell models deal with the number of cells infected. Our approach here differs from this in that it factorises the immune response into several interacting modules. This affords the opportunity for richer generative models which, using variational Bayes and Bayesian model reduction, may be inverted relatively quickly.

It is interesting that the mechanisms of resistance illustrated here tend to flatten, as opposed to fully suppress, the viral load curve. This could mean one of (or some combination of) two things. A lower load may make someone less able to transmit the infection. Alternatively, a prolonged period of having a non-negligible viral load may increase the period for which they are infectious. Each of these has important consequences in SEIR (or LIST) epidemiological models. The former favours resistance both to the infection itself, and to passing it on. The latter might be more consistent with the notion of a super-spreader, who has an insufficient viral load to cause symptoms that would prompt testing or isolation but is infectious for longer.

Clearly a major limitation of the work presented here is that it is based upon purely synthetic data. As such, this should be viewed as a proof-of-principle whose priors will require refinement as data from longitudinal studies [38] and vaccine trials become available [64]. Related to this are the temporally dense (daily) measurements assumed in the model fitting. In the presence of more precise prior beliefs derived from such studies, it should be possible to achieve posterior estimates with greater certainty and less data per individual. Another limitation is the relative coarseness of the immune model. There are many nuances that could be incorporated into a model of this sort. Relevant to the current SARS-CoV-2 pandemic is the direct influence of the virus on T-cell populations. Some stages of the SARS-CoV-2 infection are associated with a decrease in the functional CD8+ cell population [9] and lymphopenia [65]. There is precedent for thinking about targeting of lymphocytes by viral pathogens, with the most obvious example being the reduction in CD4 counts mediated by human immunodeficiency virus (HIV) [66]. However, the pattern in Covid-19 seems to be more a skew in the sorts of cells naïve T-cells differentiate into [67] – a viral effect on T-cell transition probabilities that could be explicitly parameterised using this approach.

One possible extension to this work would be to consider separate anatomical compartments. Here, the same mean-field model could be replicated for different parts of the body. This would be similar to the connection of multiple neural mass models in neuroscience – to model effective connectivity between different brain regions [68] – or to connectivity between multiple LIST models as was used to model SARS-CoV-2 propagation between parts of the United States [3]. This would allow for modelling of the transfer of virus between different anatomical sites (e.g., between the nasopharynx and the lungs) in addition to transfer of immune cells (e.g., between lymphoid organs and the lungs) [69]. Including this anatomical dimension may also be of use in modelling the effect of Covid-19 on specific body systems, including the increasingly recognised neurological manifestations, enabling measurements in the cerebrospinal fluid to inform the same parameters as serological measures. However, it is worth noting that there is little evidence so far for cerebrospinal viral load playing a direct role in neurological disease [70].

Finally, having discussed the translation of analysis tools from neuroscience to immunology, it is worth briefly considering other points of contact that could be exploited in future work. One of these is the importance of immunological processes in neurological disease. Another is the relationship between the brain and immune system in health. Regarding the former, there is an opportunity to combine neurobiological and immunological models. Recently, DCM for neuroimaging has been extended to accommodate fusion of several different imaging modalities [71]. This raises the possibility of fusing immunological and neurobiological DCMs, to model conditions – like multiple sclerosis [72,73] or autoimmune encephalitides [74] – in which the immune and nervous systems interact. Here the aim would be to generate both neurophysiological and immunological measurements. Given serial electroencephalograms and serological tests, it might be possible to measure the influence of immune system compartments on neural connectivity over time.

The relationship between the nervous and immune systems in health is another important topic. Here there is another opportunity to translate between theoretical neurobiology and immunology. A prominent approach in the former is active inference, which frames the internal dynamics of the nervous system as solving an inference problem based upon sensory data – where these inferences direct behaviour through effector systems [75]. From this perspective, it suffices to define the problem being solved. The optimal solution emerges from identification of this problem. This has been successful in accounting for a range of neurophysiological and behavioural observations [76-78]. Like the nervous system, the immune system exhibits complex internal dynamics [69], and interfaces with foreign agents via (molecular) receptors and effectors. It follows that the immune system may be amenable to similar approaches, which could help uncover the principles from which the interactions simulated above could emerge. Given extensive interfaces between neural and immune tissue – both directly [79-83] and via the neuroendocrine system [84] – it may be possible to take this further, through identifying the problem that these systems jointly solve. We hope to explore this avenue of theoretical neuroimmunology in future work.

## 6 – Conclusion

Ultimately, the simulations above are not designed to draw any specific conclusions about the nature of 'immunological dark matter', but to show how dynamic causal modelling could be applied to understand which mechanisms of resistance (i.e., attenuation of viral load) might be in play. However, one conclusion that can be drawn from this is the importance of not over-interpreting serological results. If alternative immunological phenotypes like those simulated above exist in populations affected by Covid-19, serological measures alone are likely to underestimate the effective immunity (or resistance) in the population. We have illustrated the use of a dynamic model of immune responses to supplement empirical and conceptual analyses of responses to viral pathogens or vaccines. In addition, we found that – using synthetic data – it is possible to recover the models that generate immune responses and to test hypotheses about the relationship between demographic factors and parameters of these models.

**Software note**

The simulations presented here may be reproduced and customised using the **DEM_Immune.m** demo which will be available as part of the next public release of SPM12. This academic freeware may be downloaded from https://www.fil.ion.ucl.ac.uk/spm/software/spm12/.

**Acknowledgements**

The Wellcome Centre for Human Neuroimaging is supported by core funding from Wellcome [203147/Z/16/Z]. AB is supported by a Medical Research Council doctoral studentship [D79/543369/D-OTH/170890]. KJF is a Wellcome Principle Research Fellow [088130/Z/09/Z].